\documentclass[aps,pre,twocolumn,floatfix]{revtex4}
\usepackage[pdftex]{color}
\usepackage{latexsym,hyperref,amssymb,amsmath,gensymb}
\usepackage{graphicx,dcolumn,bm,setspace}
\def\geqsim{\lower.73ex\hbox{$\sim$}\llap{\raise.4ex\hbox{$>$}}$\,$}
\def\leqsim{\lower.73ex\hbox{$\sim$}\llap{\raise.4ex\hbox{$<$}}$\,$}
\newcommand{\half}{\frac{1}{2}}

\newcommand{\bos}{\boldsymbol}

\newcommand{\beq}{\begin{equation}}
\newcommand{\eeq}{\end{equation}}
\newcommand{\bea}{\begin{eqnarray}}
\newcommand{\eea}{\end{eqnarray}}
\newcommand{\barr}{\begin{array}}
\newcommand{\earr}{\end{array}}
\newcommand{\bean}{\begin{eqnarray*}}
\newcommand{\eean}{\end{eqnarray*}}
\newcommand{\bei}{\begin{itemize}}
\newcommand{\eei}{\end{itemize}}
\newcommand{\ben}{\begin{enumeration}}
\newcommand{\een}{\end{enumeration}}
\newcommand{\nn}{\nonumber}

\newcommand{\la}{\langle}
\newcommand{\ra}{\rangle}
\newcommand{\lt}{\left}
\newcommand{\rt}{\right}
\newcommand{\ep}{\epsilon}
\definecolor{navyblue}{rgb}{.05,0,.55}

\newcommand{\tcb}[1]{\textcolor{blue}{#1}}

\begin{document}
\begin{widetext}
\begin{flushleft}
{\Large{\textbf{\textsf{Simple Model of the Transduction of
Cell-Penetrating Peptides}}}}
\end{flushleft}

\begin{flushleft}
\textsf{Kevin Cahill\hfill\today\\cahill@unm.edu}\\
\textsf{Biophysics Group,
Department of Physics \& Astronomy,
University of New Mexico, Albuquerque, NM 87131, USA}
\end{flushleft}

\pagestyle{myheadings}
\markright{Capacitor Model of Transduction}
\begin{abstract}\noindent
\textsf{ABSTRACT \quad
Cell-penetrating peptides (CPPs) 
such as HIV's \tcb{t}rans-\tcb{a}ctivating \tcb{t}ranscriptional
activator (\tcb{TAT}) and polyarginine 
rapidly pass through the plasma membranes of mammalian cells
by an unknown mechanism called transduction. 
They may be medically useful when fused to well-chosen
chains of fewer than about 35 amino acids.
I offer a simple model of transduction
in which phosphatidylserines and CPPs
effectively form
two plates of a capacitor
with a voltage sufficient
to cause the formation
of transient pores (electroporation).
The model is consistent with
experimental data 
on the transduction of oligoarginine
into mouse C\(_2\)C\(_{12}\) myoblasts
and makes three testable predictions.}
\end{abstract}

\maketitle
\end{widetext}
\section{Cell-Penetrating Peptides
\label{Cell-Penetrating Peptides}}
Polyarginine, TAT~\citep{Loewenstein1988,Pabo1988},
Penetratin, and other short, positively
charged peptides can penetrate 
the plasma membranes of live cells
and can tow along with them cargos
that greatly exceed the 600 Da restriction barrier.
They are promising therapeutic 
tools when fused to well-chosen 
sequences of fewer than about 35 amino 
acids~\citep{Tsien2004,Dowdy2005,Pugh2002,Kaelin1999,Fong2003,Cohen2002,Robbins2001,Datta2001,Hosotani2002,Hsieh2006,SnyderDowdy2004,Morano2006,Tuennemann2007}\@.
\par
TAT carries cargos across 
cell membranes with high efficiency
by at least two functionally distinct mechanisms
according to whether the cargo is big
or small~\citep{Cardoso2006}\@.
Big cargos, such as proteins
or quantum dots, enter via 
caveolae endocytosis
and macropinocytosis~\citep{Dowdy2004,Brock2007},
and relatively few escape the 
cytoplasmic vesicles in which they then
are trapped~\citep{Cardoso2006}\@.
\par
Small cargos, such as peptides 
of fewer than 30--40 amino acids, 
enter both slowly
by endocytosis and rapidly by transduction,
an unknown mechanism
that uses the membrane 
potential~\citep{Cardoso2006,Prochiantz2000,Rezsohazy2003,Wender2005}\@.
Peptides fused to TAT
enter cells
within seconds~\citep{Seelig2005}\@.
\par
Sec.~\ref{Plasma Membranes}
reviews some facts about
plasma membranes.
Sec.~\ref{The Puzzle} reviews how
the phospholipid bilayer prevents ions
from crossing the plasma membrane.
Sec.~\ref{A CPP--PS Capacitor and Electroporation}
describes a simple model of transduction 
in which CPPs on the outer leaflet
and phosphatidylserines on the inner leaflet
form a kind of capacitor with a voltage
sufficient to favor the formation
of transient pores (electroporation)\@.
Sec.~\ref{Comparison with Experiment}
shows that the model is consistent
with measurements made by
T\"{u}nnemann \textit{et al.}~\citep{Cardoso2008}
on the transduction into mouse myoblasts 
of oligoarginines and oligolysines carrying fluorophores
of 400 Da\@.
Sec.~\ref{How to Test This Model}
tells how to test three predictions of the model.
The paper ends with a short summary
in Sec.\ref{Summary}\@.

\section{Plasma Membranes
\label{Plasma Membranes}}
The plasma membrane of a mammalian cell
is a lipid bilayer that is 4 or 5 nm thick.
Of the four main phospholipids in it,
three---phosphatidylethanolamine (PE),
phosphatidylcholine (PC), and
sphingomyelin (SM)---are neutral,
and one, phosphatidylserine (PS), 
is negatively charged\@.
In live cells, PE and PS 
are mostly in the cytosolic layer,
and PC and SM in the 
outer layer~\citep{Zwaal1999,MBoC4587}\@.  
Aminophospholipid translocase (flippase) moves 
PE and PS 
to the inner layer;
floppase slowly moves all phospholipids
to the outer layer~\citep{Zwaal1999}\@.
\par
Glycolipids make up about 5\%
of the lipid molecules of the outer layer
of a mammalian plasma membrane.
Their hydrocarbon tails normally
are saturated.  Instead of a modified
phosphate group, they are decorated 
with galactose, glucose, 
GalNAc = N-acetylgalactosamine,
and other sugars.  The most complex
glycolipids---the gangliosides---have 
negatively charged 
sialic-acid (NANA) groups\@.
Incidentally, cholera toxin binds to and enters cells 
that display the G\(_{M1}\)
ganglioside.~\citep{MBoC4587}
\par
A living cell maintains an
electrostatic potential of 
between 20 and 120 mV
across its plasma membrane.
The electric field \(E\) within
the membrane points into the cell
and is huge, about 15 mV/nm or
\(1.5 \times 10^7\) V/m
if the potential difference 
is 60 mV across a membrane of 4 nm.
Conventionally, one reports membrane
potentials as the electric potential
inside the cell minus that outside,
so that here \(\Delta V = - 60\) mV\@.
Near but outside the membrane,
this electric field falls-off
exponentially \(E(r) = E \, \exp(-r/D_\ell)\)
with the ratio of the distance \(r\)
from the membrane to the Debye length \(D_\ell\),
which is of the order of a nanometer\@.
The rapid entry of TAT fused to peptides is frustrated
only by agents that destroy the
electric field \(E\)~\citep{Cardoso2006},
which applies a force \(qE\)
to a CPP of charge \(q\)\@.
\par
Most of the phospholipids 
of the outer leaflet of the plasma membrane
are neutral PCs \& SMs\@.
They vastly outnumber
the negatively charged gangliosides, 
which are a subset
of the glycolipids, which themselves amount only
to 5\% of the outer layer.
Imagine now that 
CPP-cargo molecules
are in the extra-cellular environment.
Many of them 
will be pinned down by the electric field
\(E(r)\) just outside the membrane,
their positively charged side-chains
interacting with the negative phosphate
groups of neutral
dipolar PC \& SM head groups.
(Other CPP-cargo molecules 
will stick to negatively
charged gangliosides and  
to glycosaminoglycans (GAGs) 
attached to transmembrane proteoglycans (PGs);
these slowly will be endocytosed.
Heparan-sulfate PGs are needed
for TAT-protein endocytosis~\citep{Giacca2001}\@.)
It is crucial that the 
dipolar PC \& SM head groups 
are neutral and so do not
cancel or reduce the positive electric charge
of a CPP-cargo molecule.
The net positive charge of a CPP-cargo
molecule will attract negatively charged
PSs and cause them to diffuse toward
the part of the inner leaflet directly
below the CPP-cargo molecule.
This is the starting point for the model
described in the next two sections.

\section{The Puzzle
\label{The Puzzle}}

The dielectric constant 
\(\epsilon_{\ell} \approx 2\) 
of the hydrocarbons of a lipid bilayer
is much less than that of
water \(\epsilon_w \approx 80\)\@.  
Thus, the difference \(\Delta E_{w \to \ell}\) 
in the electrostatic
energy of an ion of charge \(q\)
and effective radius \(a\) in the bilayer
and in water~\citep{Parsegian1969} is
\beq
\Delta E_{w \to \ell} = 
\frac{q^2}{8 \pi \epsilon_0 a}
\lt( \frac{1}{\epsilon_{\ell}} 
-  \frac{1}{\epsilon_w} \rt)
\label {Delta E wl}
\eeq
or 3.5 eV 
if the ion's charge is that
of the proton and its
radius is \(a = 1\) \AA\@. 
This energy barrier is far larger
than the 0.06 eV gained
when a unit charge crosses
a 60 mV phospholipid bilayer.
Thus, an ion will not cross
a cell's plasma membrane
unless a transporter or a channel 
facilitates (and regulates) its passage.
\par
In the present model of CPP transduction,
the electrostatics
of the CPP-cargo complex and 
the role of PSs on the inner leaflet
play key roles.
\par
The electrostatics of 
a cationic polypeptide such
as TAT or polyarginine are
more complex than for an ion.  
I will model
the CPP and its cargo
in water as a sphere with its 
positive charges on its surface.
The density of a protein
of mass \(M\) kDa is estimated~\citep{Craievich2004}
to be
\beq
\rho(M) = \lt( 0.8491 + 0.0873 \, e^{-M/13} \rt)
\quad \mbox{kDa/nm}^3.
\label {rho(M)}
\eeq
A CPP-cargo complex would not be expected
to fold
as densely as a natural globular protein, 
and so for it the
estimate \(\rho(M)\) is more of an upper bound.
The radius \(r\) of a putative sphere 
consisting of \(M\) kDa of CPP and cargo
then would be at least
\beq
r \gtrsim \lt(\frac{3}{4\pi} 
\frac{M}{\rho(M)}\rt)^{1/3}
\quad \mbox{nm}.
\label {r =}
\eeq
For instance,  
a CPP of \(N\)
arginines and a tiny fluorophore
cargo of 400 Da 
has a mass of \(M_N = 0.1562 N + 0.4\) kDa,
and so its radius would satisfy
\beq
r \gtrsim  \lt(\frac{3}{4\pi} 
\frac{M_N}
{\rho(M_N)}\rt)^{1/3} 
\quad \mbox{nm}
\label {r NR}
\eeq
which gives \(r = 0.75\) nm
for \(N = 8\) arginines.
The lower bounds on the radii for \(N = \) 5--12
are listed in column 2 of 
the Table~(\ref{table of energies})\@.
\par
For larger cargos
of \(N_c = 50\)--100 amino acids
of 130 Da each, the lower bounds
on the radii
range from 1.25 to 1.59 nm.
(In what follows, \( N_c \) will represent the number
of amino acids in the cargo or the mass of the cargo in
Daltons divided by 130 Da.) 
Adding another 0.8 nm for the PC/SM
head groups extends these lower bounds
on the radii
to 2.05--2.39 nm.  The diameters
of these spheres would approach or exceed
the normal 4.4 nm of the lipid bilayer.
The extra energy penalty
due to the distortion of
the plasma membrane may 
explain why CPPs can not
transduce cargos of more
than 50 amino acids~\citep{Cardoso2006}\@.
\par
If the CPP-cargo molecule were
a charged conducting sphere
of radius \(r\) and charge \(q\),
then its electrostatic energy
in water would be
\beq
E(N,N_c,q,w) = 
\frac{q^2}{8\pi \epsilon_0 \epsilon_w r}.
\label {E(N,N_c,w)}
\eeq
This term neglects the
short-distance detail of
the electric field near
the \(q/e\) positive unit charges \(e\)
of the CPP-cargo molecule.  
So a short-distance correction term 
\beq
E_{sdc}(N,N_c,q,w) = 
\frac{qe}{8\pi \epsilon_0 \epsilon_w a}
\label {Ecw}
\eeq
proportional to \(q\) 
must be added to \( E(N,N_c,q,w) \)\@.
The short distance \(a\)
is an adjustable parameter,
which should turn out to
be several \AA\ since 
the term \(E_{sdc}\) 
is a correction to be added
to \(E(N,N_c,q,w)\) and not
the entire electrostatic energy.
\par
As the CPP-cargo molecule
enters the lipid bilayer,
the phosphate groups
of the PC and SM of the
outer leaflet bind to
the positively charged
guanidinium and amine groups
of the CPP\@.  
The positive charges of the
phosphocholine groups of PC and
SM are about \(d =5\) \AA\  from
their phosphate groups~\citep{MBoC5.620}\@.
The binding of PC and SM
therefore approximately
increases the effective
radius of the charged
sphere to \(r_m \approx r + d\)\@.
The electrostatic energy
of this complex in the hydrocarbon
tails of the lipid bilayer then is
\beq
E(N,N_c,q,\ell) \approx
\frac{q^2}{8\pi \epsilon_0 \epsilon_{\ell} (r+d)}
\label {E(N,N_c,m)}
\eeq
apart from a correction factor
\beq
E_{sdc}(N,N_c,q,\ell) = 
\frac{qe}{8\pi \epsilon_0 \epsilon_{\ell} a}
\label {Ecl}
\eeq
similar to (\ref{Ecw})\@.
\par
Apart from correction terms,
the electrostatic energy penalty when the
CPP-cargo molecule enters the lipid bilayer
from water as a CPP-cargo-PC/SM
complex is the difference
\bea
\Delta E_{w, \ell}^0(N,N_c,q) 
& \approx & E(N,N_c,q,\ell) - E(N,N_c,q,w) 
\label {dE NN_cq} \\
& \approx & \frac{q^2}
{8\pi \epsilon_0 \epsilon_{\ell} (r+d)}
\lt( 1 - \frac{r+d}{r}
\frac{\epsilon_{\ell}}{\epsilon_w}
\rt). \nn
\eea
\par
Because the thickness \(t = 4.4\) nm
of the lipid bilayer
is only a few times that of the
CPP-cargo-PC/SM complex, we also
must include the 
Parsegian correction~\citep{Parsegian1969}
\beq
\Delta E_P = - 
\frac{q^2}{4\pi \epsilon_0 \epsilon_{\ell} t}
\ln\lt(\frac{2 \epsilon_w}{\epsilon_w + \epsilon_{\ell}}
\rt).
\label {E_P}
\eeq
The sum of the water-to-lipid energy (\ref{dE NN_cq})
and Parsegian's correction (\ref{E_P}) is
\beq
\Delta E_{w, \ell}(N,N_c,q) = 
\Delta E_{w, \ell}^0(N,N_c,q) + \Delta E_P .
\label {dEwlP}
\eeq
The energy \(\Delta E_{w, \ell}(N,N_c,q)\)
is listed in column~3 of 
Table~(\ref{table of energies})
for a CPP of \(N = \) 5--12 arginines towing
a fluorophore cargo of 400 Da
with \(d = 0.5\) nm.
\par
The short-distance correction terms 
augment this penalty by 
\bea
\Delta E_{sdc}(N,N_c,q) & = & E_{sdc}(N,N_c,q,\ell)
- E_{sdc}(N,N_c,q,w) \nn\\
& = & \frac{qe}{8\pi \epsilon_0 \epsilon_{\ell} a}
\lt( 1 - \frac{\epsilon_{\ell}}{\epsilon_w} \rt)
\label {dEsdc}
\eea
and do not require Parsegian's correction
because they are short-distance effects.
This short-distance correction \(\Delta E_{sdc}\)
is listed in column~4 of 
Table~(\ref{table of energies})
for a CPP of \(N = \) 5--12 arginines
and a representative value of 
\(a = 4.5\) \AA\ for the short-distance parameter.
\par
The net electrostatic energy penalty when the
CPP-cargo molecule enters the lipid bilayer
from water as a CPP-cargo-PC/SM
complex is then the sum of 
(\ref{dE NN_cq}, \ref{E_P}, \& \ref{dEsdc})
\beq
\Delta E_{w \to \ell} =
\Delta E_{w,\ell}^0 + \Delta E_P + \Delta E_{sdc}.
\label {full electrostatic cost}
\eeq
\par
A CPP of 8 arginines carrying
a fluorophore of 400 Da has a radius
\(r\) of 0.75 nm, and with \(a = 4.5\) \AA,
the change (\ref{full electrostatic cost})
in its electrostatic energy
on going from water to lipid is 
\beq
\Delta E_{w \to \ell}(8,3,8e) \approx
16.9 \; \mbox{eV}.
\label {dE(9,15,9e)}
\eeq
This energy barrier is 35
times bigger than the energy
0.48 eV that
it gains by crossing
a potential difference of 60 mV\@.
So how and why does it cross?
\par
\begin{table}
\caption{\label{tab:table of energies}
The radius \(r\) of a CPP-cargo molecule
of \(N\) arginines and a cargo of 400 Da,
its change in electrostatic energy 
\(\Delta E_{w,\ell}\)
when transferred from water to hydrocarbon
(\ref{dEwlP}), and
the short-distance correction \(\Delta E_{SDC}\)
(\ref{dEsdc})\@.
Distances are in nm and 
energies in electron-Volts.}
\label {table of energies}
\begin{ruledtabular}
\begin{tabular}{||c|c|c|c||} 
\, \(N\) \quad & \(r\) \quad
& \(\Delta E_{w, \ell}\) \quad  
& \(\Delta E_{sdc}\) \quad  \\ \hline 
5 & 0.67 & 4.61 & 3.90 \\ \hline
6 & 0.70 & 6.39 & 4.68 \\ \hline
7 & 0.73 & 8.41 & 5.46 \\ \hline
8 & 0.75 & 10.64 & 6.24 \\ \hline
9 & 0.78 & 13.06 & 7.02 \\ \hline
10 & 0.80 & 15.68 & 7.80 \\ \hline
11 & 0.82 & 18.48 & 8.58 \\ \hline
12 & 0.84 & 21.44 & 9.36 \\ \hline
\end{tabular}
\end{ruledtabular}
\end{table}
\section{A  CPP--PS Capacitor and Electroporation
\label{A  CPP--PS Capacitor and Electroporation}}

My answer lies in
the second part of the present 
model---the phosphatidylserines (PSs).
They comprise some 8--18\%
of the inner leaflet by weight~\citep{MBoC5.624}\@.
The PSs diffuse laterally within
that leaflet with a diffusion
constant 
\(D \approx 10^{-8}\) 
cm\(^2\)/sec~\citep{MBoC5.622}
and so within one second 
spread to an area of 12 \(\mu\)m\(^2\),
which is a significant fraction
of the surface area of a eukaryotic cell.
Attracted by the positive charge
of two or more CPPs, several PSs 
cluster in the inner leaflet
below the CPPs forming with them
a kind of capacitor.  In what follows,
I will focus on the PSs because they
are constrained to lie in the 
two-dimensional inner
leaflet while other anions drift
in the three-dimensional cytosol.
\par
After oligoarginine CPPs are added to the
extra-cellular fluid medium, the arginines
of some of them will
bind to the phosphate groups of the
phosphatidylcholine (PC) and
sphingomyelin (SM) phospholipids 
in the outer leaflet.  
The electrostatic potential
of an isolated CPP may not
be strong enough to form a pore
because it is reduced
at a point \( \bos{r} \) in the lipid
by the mean relative permittivity 
\( \bar \epsilon = (\epsilon_w + \epsilon_\ell)/2 \)
to
\(
V(\bos{r}) = q/(4\pi \ep_0 \bar \epsilon r)
\)
and further reduced by counterions
and image charges.
But if a second CPP or a few CPPs
should land near the first CPP
and bind to PCs or SMs,
then transduction 
becomes possible in the present model.
\par
A CPP of \( N \) arginines can form
an \( \alpha \)-helix of length
\( L \approx 0.16 \, N \) nm (or a \( \beta \)-strand
of length \( 0.34 \, N \) nm)\@.
If two \( R^N \) CPPs 
lie a distance \(d\) apart bound to phosphate groups
of the outer leaflet, then the charge \( 2 N e \)
of the two CPPs is spread over approximately 
\(L \, d\), giving a surface-charge density of
\( \sigma = 2 \, N \, e/ ( d \, L) \)\@.
The electric field of the two oligoarginines
would attract \(2 N\) PSs to the part of
the inner leaflet below the
two R\(^N\)s forming a kind of capacitor.
If the voltage drop across this
capacitor exceeds the threshold for
the formation of the pre-pores or pores
of electroporation, then the CPP and its
cargo may enter the cell.  This threshold
is between roughly \(- 250\) and \(\- 550\) 
mV, depending upon the duration
of the potential~\citep{Chernomordik2001}\@.
\par
Due to its bidentate guanidinium group,
arginine binds to phosphate groups
better than lysine~\citep{Wender2002,Wender2005}\@.
Thus, oligoarginines would be more likely
than oligolysines to bind to the
phospholipids of the outer leaflet 
forming transiently stable capacitors 
with the PSs of the inner leaflet.
\par
CPPs with \(N = 12\) arginines can bind
to more than one phosphate group, and so
they can bind to more than one phospholipid
of the outer leaflet.  The number of such
binding sites drops with the number of Rs.
In the ``picket-fence model," transmembrane proteins 
bind to actin and restrict the motion 
of outer-leaflet proteins 
and lipids~\citep{Dlidke2008}; so 
I assume that oligoarginines
with \(N > 8\) remain approximately stable
for a few hundred ms.  
\par
To include the effects of entropy 
and of the geometry of the oligoarginines (R\(^N\)s)
and the phosphatidylserines (PSs),
I wrote and ran a Monte Carlo program
(included as supplementary material)\@.
In this code, I assumed that two
R\(^N\)s were parallel \( \alpha \)-helices
a distance \(d\) apart bound to phosphate
groups of the outer leaflet.
The PSs were allowed to move
in two dimensions in the inner leaflet
at sites \(\bm{r}_k\) for \(k = 1, \dots 2N\)\@.
The \(2 N \) positive charges
of the two R\(^N\)s are taken to lie fixed
at the points
\beq
\bos{r_{j \, CPP}^{\pm}} = 
( \pm \, d/2, \,\,0.16 \, (N/2 - j), \,\, t )
\label {r_cpp}
\eeq
in which the \(\pm\)-signs label
the two R\(^N\)s, and \(j\) labels
the Rs of the R\(^N\)s ( \( 0 \le j \le N \))\@. 
The electrostatic energy of the PSs then is
approximately
\beq
E = 
\frac{1}{4\pi \epsilon_0 \bar \epsilon}
\lt(
\sum_{i = 2}^{N_{PS}}\sum_{k=1}^{i-1}
\frac{e^2}{|\bm{r}_i - \bm{r}_k |} 
 - \sum_{k = 1}^{N_{PS}} \sum_{j = 1}^N
\frac{qe}{
|\bm{r}_i - \bm{r}_{j \, CPP}^{\pm}|}
\rt)
\label {E}
\eeq
in which \( \bar \epsilon \)
is the mean permittivity
\beq
\bar \epsilon = \half \lt( \epsilon_w
+ \epsilon_\ell \rt).
\label {bareps}
\eeq
\par
The Monte Carlo code used a simple
Metropolis step in which the \(x\)-\(y\) coordinates
of all the PSs were randomly varied,
each by less than \(\pm 0.025 \) nm.
The code accepted any change in the
positions of the PSs that lowered 
their energy \(E\) as given by (\ref{E})
and accepted any change 
that raised their energy by \(\Delta E\) 
conditionally with probability
\beq
P = e^{- \Delta E /(kT)}
\label {P}
\eeq
in which \(k\) is Boltzmann's constant
and \(T\) is 37 Celsius.
The simulation took place 
in a two-dimensional box
with side \( L = \sqrt{N/0.13} \) nm,
and periodic boundary conditions were
enforced to keep the number density
of PSs constant at 13\%
within the box.
\par
Each simulation started from a random
configuration of PSs on the inner leaflet and
ran for at least 10 million Metropolis steps.
After at least 2 million steps for thermalization,
measurements of the positions of the PSs
and of the electrostatic potential 
were taken every
1000 steps, and the positions of the PSs
were recorded every 80,000 steps.
The distnace \(d\) separating the two
CPPs was taken to be \( d =2 \), 3, and 4 nm.
(The simulations for \(d = 3\) used ten times
as many steps.)
\par
\begin{table}
\caption{\label{tab:table of voltages}
The voltage differences \( \Delta V \) (mV)
across the plasma membrane due to 
two R\(^N\) oligoarginines on the outer leaflet
separated by a distance \(d\) nm
and a flock of \(2 N \) phosphatidylserines
fluttering below them on the inner leaflet.}
\label {table of voltages}
\begin{ruledtabular}
\begin{tabular}{||c|c|c|c||} 
\, \(N\) 
& \(\Delta V\) \quad \(d = 2\) nm
& \(\Delta V\) \quad \(d = 3\) nm 
& \(\Delta V\) \quad \(d = 4\) nm  \\ \hline 
5 &  \(-342\)     & \(-229\)  & \(-173\) \\ \hline
6 &  \(-398\)     & \(-266\)  & \(-198\) \\ \hline
7 &  \(-449\)     & \(-300\)  & \(-223\) \\ \hline
8 &  \(-504\)     & \(-335\)  & \(-248\) \\ \hline
9 &  \(-553\)     & \(-369\)  & \(-273\)    \\ \hline
10 & \(-601\)     & \(-401\)  & \(-297\)    \\ \hline
11 & \(-646\)     & \(-433\)  & \(-319\)    \\ \hline
12 & \(-687\)     & \(-465\)  & \(-340\)    \\ \hline
\end{tabular}
\end{ruledtabular}
\end{table}

\par
The voltages were measured at the
points \( ( 0, 0, t ) \)
and \( ( 0, 0, -0.1 ) \) nm.
The offset by 0.1 nm was to avoid
the chance arrival of a PS at the point
\( ( 0, 0, 0 ) \)\@.
The voltage differences
\( \Delta V = V(0, 0, -0.1) - V( 0, 0, t ) \)
are displayed in the 
Table~\ref{tab:table of voltages}\@.
At all three separations,
CPPs with 9 or more Rs produce
voltages that exceed the threshold
( \(\sim 250\)--\(\sim 550\) mV )
for the formation of pre-pores or 
pores~\citep{Chernomordik2001}\@.
CPPs with 10 or more Rs may exceed
the threshold for irreversible electroporation.
\par
The mean value of the distance
of the PSs from the point \( ( 0,0,0) \)
varied between \( \la r \ra = 11.8 \) nm
for two R\(^5\)s separated by 2 nm
and \( \la r \ra = 27.5 \) nm for
two R\(^{12}\)s separated by 4 nm.
\par
In the present model,
a CPP-PS capacitor increases the
membrane potential sufficiently
to cross the threshold for
electroporation, at least for
R\(^N\)s with enough arginines.
In its use of an electric field
and of the binding of the CPPs
to the phosphate groups of the phospholipids
of the outer leaflet,
the model has something in common with
the adaptive-translocation model 
of Rothbard, Jessop, and Wender~\citep{Wender2005};
in its use of 
neutral dipolar PC \& SM head groups
it is somewhat similar to the work of 
Herce and Garcia~\citep{Garcia2007}
and of Tang, Waring, and Hong~\citep{Hong2007}\@.
The key distinctive feature
of the present model is the 
role of PSs 
in forming with the CPPs
a CPP-PS capacitor with a voltage
high enough to cause electropration.

\section{Comparison with Experiment
\label{Comparison with Experiment}}

T\"unnemann \textit{et al.}~\citep{Cardoso2008}
measured the ability of
the L- and D-isoforms of oligolysine and of
oligoarginine to carry fluorophores 
of \(\sim\!\!400\) Da into live cells.
They found that oligoarginines transduced 
the fluorophores much better than oligolysines,
that more arginines meant faster transduction,
with L-R9 and L-R10 doing better than 
their shorter counterparts, and that
the D-isoforms worked better than the L-isoforms,
probably because of their greater 
resistance to proteases.
\par
The present model is consistent
with these experimental facts and
explains them as follows:
The oligoarginines crossed cell membranes
more easily than the oligolysines
because they were better able to bind to
the phosphate groups of the PCs and
SMs in the outer leaflet; the
oligolysines were not able to
form a stable upper plate of
the CPP-PS capacitor.
CPPs with more arginines were
transduced more rapidly because
with more arginines they could
bind to more PCs and SMs and because
their higher charge led to a higher
trans-membrane potential. 
The D-isoforms worked better than the L-isoforms
because the capacitor mechanism
is insensitive to the chirality
of the amino acids and because
proteases were less able to cut them.
\par
This consistency of the capacitor model
lends it some plausibility.
The simplicity of the model suggests
that it may apply to other
oligocationic CPPs.  But evolution
finds what works, not what fits neatly
into a model, and so other CPPs with
different cargos may enter different
cells by different mechanisms.

\section{How to Test This Model
\label{How to Test This Model}}
One way to test the model
would be to compare the 
rates of polyarginine 
transduction 
in wild-type cells and in those that
have little or no phosphatidylserine (PS) 
in their plasma membranes.
If PS plays a key role as in the model
of this paper,
then the transduction of
polyarginine fused to a cargo
of less than 30 amino acids
should be faster in the wild-type cells
than in those without PS in their
plasma membranes.
Mammalian cell lines that are
deficient in the synthesis
of phosphatidylserine do 
exist~\citep{Frazier1986,Nishijima1996,Kuge1998,Vance2000,Vance2004},
but they appear to have normal levels
of PS in their 
plasma membranes~\citep{Vance2004}---presumably 
due to a lower rate of PS 
degradation~\citep{Vance2008}\@.
\par
Another test 
would be to construct artificial asymmetric 
bilayers~\citep{Schlue1993,Bezrukov1998,Ligler2007} 
with and without PS
on the ``cytosolic'' side and to compare
the rates of CPP-cargo transduction.
If the present model is right,
then the rate of transduction
should be higher through membranes
with PS on the cytosolic side than
through membranes with no PS or with
PS on both sides.
\par
If CPPs do enter cells via molecular
electroporation, then 
it may be possible 
to observe the formation
of the transient pores by 
detecting changes in the conductance
of the membrane~\citep{Chernomordik2001}\@.
Such measurements would be
a third test of the model 
and would let one 
determine both
whether CPP-transduction
is related to the presence
of PS on the cytosolic side of the
membrane and whether it
proceeds via molecular
electroporation.

\section{Summary
\label{Summary}}

Cell-penetrating peptides can carry into cells
cargos with molecular weights 
of an order of magnitude greater than
the nominal limit 500 of the 
``rule of 5''~\citep{Lipinski1997}\@.
Section~\ref{A CPP--PS Capacitor and Electroporation}
describes a model in which phosphatidylserines (PSs)
play key roles
in the transduction of CPP-cargo molecules.
In this model, the positively charged CPP
on the outer leaflet and the negatively charged
PSs on the inner leaflet form a kind 
of capacitor with a trans-membrane potential
sufficient for electroporation
(-250 to -550 mV)\@.
The model is consistent with
experimental data~\citep{Cardoso2008}
on the transduction
of the L- and D-isomers
of oligoarginine and oligolysine
CPPs into mouse myoblasts.
\par
The model predicts that
mammalian cells that lack
phosphatidylserine in their
plasma membranes
transduce oligocations poorly,
that artificial asymmetric 
bilayers with PS
on the cytosolic side 
transduce oligocations better than
ones without PS, and that
transduction is accompanied
by brief changes in the conductance
of the membrane.
\begin{acknowledgments}
I am grateful to Gisela T\"{u}nnemann
for sending me some of the data 
from her group and for helpful e-mail,
to John Connor and Karlheinz Hilber for 
explaining the status of measurements
of the membrane potential of mouse myoblast cells,
to Paul Robbins for sending me some
of his images, and to James Thomas
for helpful discussions.
Thanks also to H.~Berg, S.~Bezrukov, H.~Bryant, 
P.~Cahill, D.~Cromer, T.~Hess, D.~Lidke, K.~Lidke, S.~Koch,
V.~Madhok, A.~Parsegian, S.~Prasad, B.~B. Rivers, and
K.~Thickman
for useful conversations, and to
K.~Dill, S.~Dowdy, S.~Henry, K.~Hilber, A.~Pasquinelli, 
B.~Salzberg, D.~Sergatskov, L.~Sillerud, 
A.~Strongin, R.~Tsien, J.~Vance,
and A.~Ziegler 
for helpful e-mail.
\end{acknowledgments}
\bibliography{bio,physics}
\end{document}